\documentclass[aps,prl,twocolumn,superscriptaddress]{revtex4}
\usepackage{graphicx}
\input epsf
\usepackage{amsmath,amssymb}
\begin{document}
\title{Selenization of V$_2$O$_5$/WO$_3$ Bilayers  for Tuned Optoelectronic Response of WSe$_2$ Films.}
\author{Abhishek Bajgain}
\affiliation{DOD Center of Excellence for Advanced Electro-Photonics with 2D Materials, Morgan State University, Baltimore, MD, 21251, USA}
\author{Santu Prasad Jana}
\author{ Alexander Samokhvalov}
\affiliation{DOD Center of Excellence for Advanced Electro-Photonics with 2D Materials, Morgan State University, Baltimore, MD, 21251, USA}
\author{Thomas Parker}
\affiliation{DEVCOM Army Research Laboratory, Aberdeen Proving Ground, Aberdeen, MD, 21005, USA}
\author{John Derek Demaree}
\affiliation{DEVCOM Army Research Laboratory, Aberdeen Proving Ground, Aberdeen, MD, 21005, USA}
\author{Ramesh C. Budhani}
\email {ramesh.budhani@morgan.edu}
\affiliation{DOD Center of Excellence for Advanced Electro-Photonics with 2D Materials, Morgan State University, Baltimore, MD, 21251, USA}
\begin{abstract}
Scalable and controlled doping of two-dimensional transition metal dichalcogenides is essential for tuning their electronic and optoelectronic properties. In this work, we demonstrate a robust approach for substitution of vanadium in tungsten diselenide (WSe$_2$) via the selenization of pre-deposited V$_2$O$_5$/WO$_3$ thin films. By adjusting the thickness of the vanadium oxide layer, the V concentration in W$_{1-x}$V$_x$Se$_2$ is systematically varied. Electrical measurements on field-effect transistors reveal a substantial enhancement in hole conduction, with drain current increasing by nearly three orders of magnitude compared to undoped WSe$_2$. Temperature-dependent electrical resistivity indicates a clear insulator-to-metal transition with increasing V content, likely due to band structure modifications. Concurrently, the photoconductive gain decreases, suggesting enhanced recombination and charge screening effects. These results establish vanadium doping via selenization of V$_2$O$_5$/WO$_3$ films as a scalable strategy for modulating the transport and photoresponse of WSe$_2$, offering promising implications for wafer-scale optoelectronic device integration.
\end{abstract}
\maketitle

Semiconducting two-dimensional transition metal dichalcogenides (TMDs) \cite{TMDs1} have gained significant interest for their potential use in energy-efficient optoelectronics, nanoelectronics, memory storage, catalysis, and surface-mounted sensing \cite{TMDs2,TMDs3,TMDs4,TMDs5}. Their atomically thin structure, natural abundance, environmental stability, mechanical flexibility, tunable bandgap, high mobility of charge carriers (compared to organic semiconductors), and excellent electrostatic gate control make them highly promising for these applications. While graphene remains the most extensively studied 2D material, its gapless nature limits its applicability \cite{graphene1,graphene2}. In contrast, 2D-TMDs such as MoS$_2$ \cite{optoelectronic1} and WS$_2$ \cite{optoelectronic2} have gained considerable attention for optoelectronic applications. However, there is a broader range of TMDs that hold great potential for energy-related applications, necessitating further research. Among these, tungsten diselenide (WSe$_2$) is particularly promising, featuring a direct bandgap of approximately 1.65 eV in monolayers and an indirect bandgap of around 1.2 eV in bulk form \cite{bandgap1,bandgap2}. Studies have demonstrated its potential in transistors \cite{application1}, sensors \cite{application2,application3}, broadband photodetectors \cite{application4}, and flexible electronics \cite{application5}. For practical integration of these 2D materials into electronic and optoelectronic devices, it is crucial to grow large single-crystal domains free of grain boundaries to ensure consistent electronic properties and high device yield. Several techniques have been explored to achieve large-area WSe$_2$ films, including chemical vapor deposition (CVD) \cite{method1}, atomic layer deposition \cite{method2}, molecular beam epitaxy \cite{method3}, van der Waals epitaxy \cite{method4}, metal vapor-assisted CVD \cite{method5}, atmospheric pressure CVD \cite{method6}, halide-assisted atmospheric-pressure CVD \cite{method7}, reverse flow CVD \cite{method8}, and molten-salt assisted CVD \cite{method9}. Despite these advancements, several challenges remain, including achieving uniform nucleation, controlling grain boundaries, defects, and strain, preventing contamination, and ensuring scalability. Additionally, high-temperature requirements can limit substrate compatibility, while optimizing reaction parameters across different substrates and finding suitable precursors remain significant concerns. 
 \par Besides scalable growth, doping remains a major bottleneck in utilizing 2D TMDs for next-generation optoelectronic devices. Several doping strategies have been explored for WSe$_2$ films, including surface charge transfer doping using molecular dopants \cite{doping1} and metal nanoparticles \cite{doping2}, electrostatic doping \cite{doping3,doping4}, intercalation doping \cite{doping5,doping6}, substitutional doping \cite{doping7,doping8}, and surface treatment techniques such as electron beam irradiation \cite{doping9}, photolithography-induced doping \cite{doping10}, vapor XeF$_2$ treatment \cite{doping11}, contact interface engineering \cite{doping12,doping13}, and ferroelectric proximity doping \cite{doping14}. While each strategy offers unique advantages, challenges persist, including achieving uniform dopant distribution, maintaining structural integrity, minimizing defect formation, preventing surface contamination, and avoiding over-etching (in surface treatment methods). Additionally, although prototype devices using vanadium dopants \cite{doping15,doping16,doping17} have been demonstrated, achieving uniform impurity distribution and precise control over dopant density at a large scale remain a challenge. Therefore, further research is essential to develop scalable doping techniques for WSe$_2$ while optimizing its optoelectronic properties for practical device applications.
\par In this paper, we report on a scalable and facile method for controlled vanadium doping of WSe$_2$ via the selenization of pre-deposited V$_2$O$_5$/WO$_3$ films. The concentration of vanadium in W$_{1-x}$V$_x$Se$_2$ is varied by using vanadium oxide (V$_2$O$_5$) films of different thicknesses, where vanadium substitutes for tungsten atoms in the lattice. The field-effect transistor (FET) performance and photoconductive response of vanadium-substituted W$_{0.83}$V$_{0.17}$Se$_2$ and W$_{0.68}$V$_{0.32}$Se$_2$ devices are compared with those of pure, undoped WSe$_2$, all measured under vacuum conditions. Both vanadium-doped devices exhibit an increase in drain current by approximately three orders of magnitude compared to pure WSe$_2$. Detailed temperature-dependent transfer characteristics reveal a transition from semiconducting to metallic behavior upon doping. This transition is attributed to dopant-induced modifications in the band structure, likely due to the presence of a metallic vanadium selenide (VSe$_2$) phase. This interpretation is further supported by the observed reduction in gate-dependent maximum photoconductive gain from 30\% in pure WSe$_2$ FETs to 8\% in W$_{0.83}$V$_{0.17}$Se$_2$ , and to an almost negligible value in W$_{0.68}$V$_{0.32}$Se$_2$  devices. This study demonstrates that selenization of pre-deposited  V$_2$O$_5$/WO$_3$ films offers a promising and scalable approach for doping WSe$_2$, while also providing valuable insights into its optoelectronic properties for practical device applications.
 
  \begin{figure*}
\centering
\includegraphics[width=16cm]{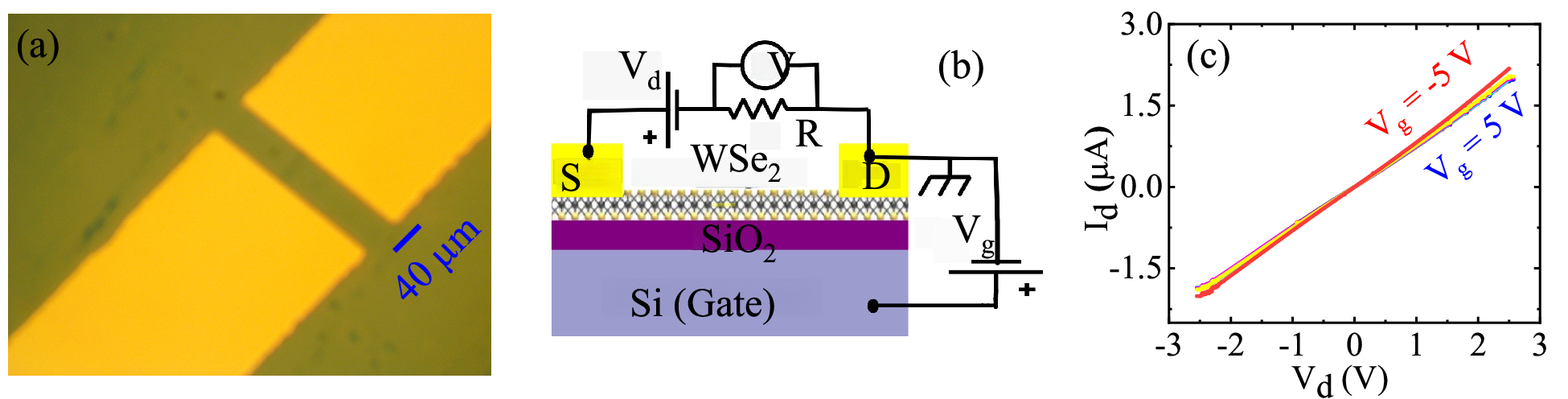}
\caption{(a) Optical image of a WSe$_2$ FET on SiO$_2$(300 nm)/Si  substrate with two probe silver contacts made by the wire masking method. (b) The schematic drawing of WSe$_2$ FET with electical measurement circuit. (c) $I_{\rm d} - V_{\rm d}$  characteristics of WSe$_2$ FET at different $V_{\rm g}$, illustrating the ohmic contacts.}
\label{fig:FIG1}
\end{figure*}      
 Highly doped Si wafers with a 300 nm thermally grown SiO$_2$ layer, cleaned using acetone and isopropyl alcohol are used as substrates with the back-gate configuration. First, vanadium oxide films of varying thicknesses were deposited on Si/SiO$_2$ substrates via thermal evaporation, followed by the deposition of WO$_3$ films. Subsequent selenization in an atmospheric-pressure CVD reactor yielded high-quality, uniform W$_{1-x}$V$_x$Se$_2$ films. After growth, films were characterized by Raman spectroscopy, Rutherford backscattering spectrometry (RBS) and X-ray photoelectron spectroscopy (XPS). The details of the thermal evaporation of V$_2$O$_5$/WO$_3$ films, their selenization process, and the associated characterization techniques are provided in the Supplementary Material.
\par For electrical measurements, source-drain contacts were fabricated using a mechanical masking technique with a 40 µm diameter copper wire, aligned under an optical microscope. A 50 nm-thick silver film was then deposited via thermal evaporation to establish the contacts. 
\par Figure \ref{fig:FIG1}(a) shows an optical micrograph of the fabricated WSe$_2$ device with source-drain contacts. Two probe conductance measurements were performed down to 200 K, following the configuration illustrated in Fig. \ref{fig:FIG1}(b). A drain-source voltage bias ($V_{\rm d}$) was applied using a programmable voltage source, while the drain current ($I_{\rm d}$) was determined by measuring the voltage drop across a series resistor (R = 119 Ω). Since the device resistance was significantly higher than 119 Ω, the series resistor ensured an accurate measurement of $I_{\rm d}$ for all values of $V_{\rm d}$. The linear two-probe current-voltage (I-V) characteristics confirm the ohmic contacts, as shown in Fig. \ref{fig:FIG1}(c). The measurement was done under a vacuum of $<10^{\rm -4}$ torr. The sample temperature was regulated using liquid nitrogen with a heater and a temperature sensor. The photoconductive response of both doped and undoped WSe$_2$ films was measured using a custom-built optical setup equipped with 532 nm (green) diode laser, featuring a beam diameter of approximately 3 mm and a maximum power of 100 mW. To apply bias and record the photo-response, an adjustable power supply (Philips PM 2821) was used to bias the device, while a nanovoltmeter (Keithley 2182A) was employed for precise voltage measurements.

\begin{figure}
\centering
\includegraphics[width=8cm]{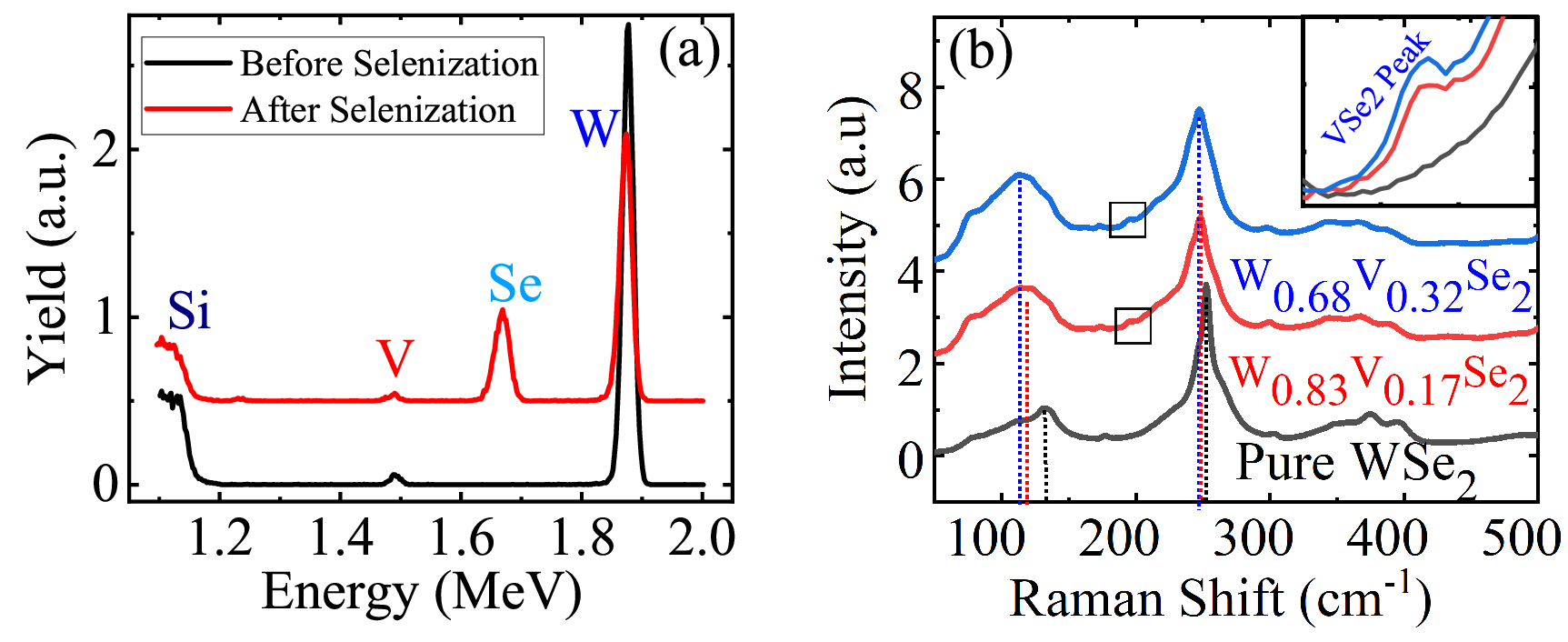}
\caption{(a) Rutherford Backscattering spectra before and after the selenization of W$_{0.83}$V$_{0.17}$Se$_2$ film. (b) Raman spectra of pristine, W$_{0.83}$V$_{0.17}$Se$_2$, and W$_{0.68}$V$_{0.32}$Se$_2$  films.}
\label{fig:FIG2}
\end{figure}
 \par Figure \ref{fig:FIG2}(a) shows the RBS spectra of the V$_2$O$_5$/WO$_3$ films before and after the selenization process. In the pre selenization state (black spectrum), distinct backscattering yields corresponding to W, V, and Si atoms in the sample are visible. After selenization (red spectrum), a strong Se peak appears, confirming the successful conversion of V$_2$O$_5$/WO$_3$ into W$_{1-x}$V$_x$Se$_2$. The value of x=0.17 corresponds to samples with pre-selenization WO$_3$ and V$_2$O$_5$ thicknesses of 11 nm and 1.9 nm, respectively, resulting in a WSe$_2$ film approximately 15 nm thick after selenization. For another sample with 9.1 nm WO$_3$ and 3.8 nm V$_2$O$_5$, the resulting WSe$_2$ film was about 16 nm thick, and the calculated value of x was 0.32. 
\par The spectrum shown in black in Fig. \ref{fig:FIG2}(b) presents the Raman data of pure WSe$_2$, where the $A _{\rm 1g} $ mode is observed at 257 $cm ^{-1} $ with a significantly reduced intensity, while the $E^{\rm 1}_{\rm 2g} $ mode appears at 253 $cm ^{-1} $. The redshift and intensity reduction of the $A _{\rm 1g} $ mode, along with the blueshift of the $E^{\rm 1}_{\rm 2g} $ mode, can be attributed to the enhanced interlayer van der Waals interactions in the multilayer WSe$_2$ compared to the monolayer, where the corresponding peaks appears at  259 $cm ^{-1} $ and 248 $cm ^{-1} $ respectively \cite{Raman1}. The red and blue spectra correspond to the W$_{0.83}$V$_{0.17}$Se$_2$ and  W$_{0.68}$V$_{0.32}$Se$_2$ samples, where the $E^{\rm 1}_{\rm 2g} $ mode exhibits a redshift of 5 $cm ^{-1}$ and 6 $cm ^{-1}$, respectively, with an increase in the full width at half maximum by 7.08 $cm ^{-1} $ and 8.87 $cm ^{-1} $ compared to pristine WSe$_2$. The redshift of the $E^{\rm 1}_{\rm 2g} $ mode is generally associated with phonon softening, which can result from local lattice strain and charge transfer effects induced by vanadium atoms substituting W atoms within the WSe$_2$ lattice. Additionally, the observed broadening of the $E^{\rm 1}_{\rm 2g} $ peak suggests enhanced phonon scattering, likely due to increased disorder and defect density introduced by vanadium doping.  The inset shows a zoomed-in portion of the Raman spectra, revealing a distinct peak near 197 $cm ^{-1} $ in the W$_{0.83}$V$_{0.17}$Se$_2$ and  W$_{0.68}$V$_{0.32}$Se$_2$ samples. This peak most likely corresponds to the $A _{\rm 1g} $ mode of the 1T-VSe$_2$ phase, as reported in the literature \cite{VSe2}. Its presence suggests that at higher vanadium doping levels, a secondary metallic 1T-VSe$_2$ phase is likely to form within the WSe$_2$ matrix. The presence of vanadium, selenium, and tungsten was also confirmed by high-resolution XPS spectra, as discussed in the supplementary material.
\begin{figure*}
\centering
\includegraphics[width=16cm]{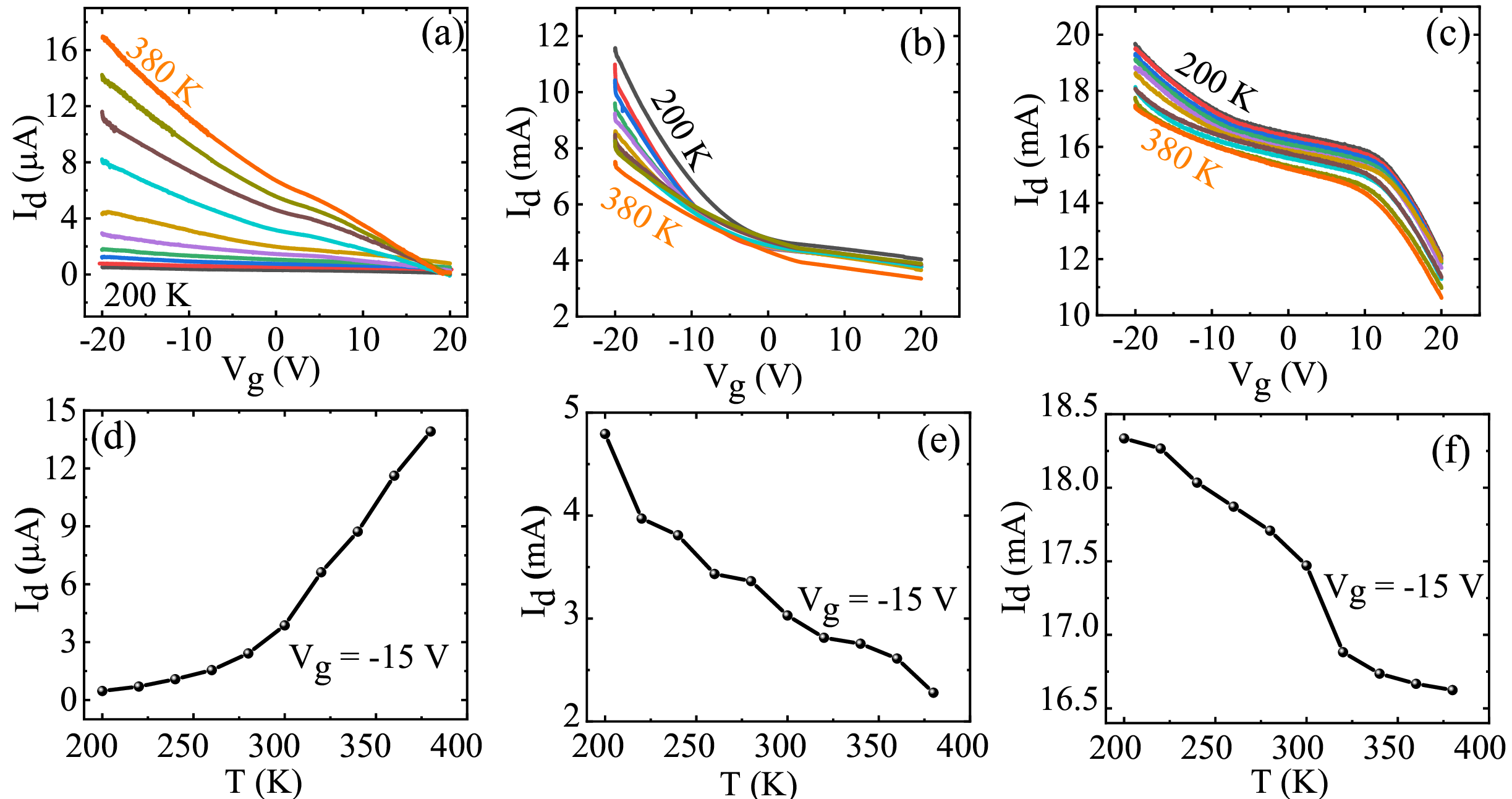}
\caption{Temperature-dependent $I_{\rm d}$ - $V_{\rm g}$ characteristics of (a) pristine, (b) W$_{0.83}$V$_{0.17}$Se$_2$, (c) W$_{0.68}$V$_{0.32}$Se$_2$ FETs at $V_{\rm d}$ = 2 V in the dark. (d), (e), and (f) show the variation of $I_{\rm d}$ with temperature at $V_{\rm g}$= -15 V extracted 
from the curves in (a), (b), and (c) respectively.}
\label{fig:FIG3}
\end{figure*}
\par Figure \ref{fig:FIG3}(a) shows the temperature-dependent transfer characteristics, i.e. $I_{\rm d}$ Vs $V_{\rm g}$ curves of undoped WSe$_2$ FETs at a fixed drain voltage $V_{\rm d}$ =  2 V, within the temperature range 200 to 380 K. The increase in $I_{\rm d}$, as shown in Fig. \ref{fig:FIG3}(a), with progressively negative gate voltage indicates that these FETs exhibit a p-type behaviour. This p-doping can be attributed to the selenium vacancies and other p-type impurities present in WSe$_2$ crystals. This leads to the pinning of Fermi energy close to the WSe$_2$ valence band maximum (VBM), facilitating efficient hole injection \cite{reason1}. However, the most remarkable observation is the substantial increase in drain current by approximately three orders of magnitude in vanadium-doped devices as shows in Figs. \ref{fig:FIG3}(b) and (c), compared to their undoped counterparts. Furthermore, the enhancement of $I_{\rm d}$ in the W$_{0.68}$V$_{0.32}$Se$_2$ sample relative to the W$_{0.83}$V$_{0.17}$Se$_2$ sample suggests that hole conduction becomes more dominant with the increasing vanadium concentration. The enhancement in hole transport can be attributed to the substitutional incorporation of vanadium atoms at tungsten sites. Vanadium, being an electron-poor dopant compared to tungsten, introduces discrete defect states near the VBM, effectively reducing the bandgap. This bandgap modulation and the resultant shift in the projected density of states have been confirmed by scanning tunneling microscopy  experiments and density functional theory simulations \cite{doping15}. Fig. \ref{fig:FIG3}(d) presents the temperature dependence of $I_{\rm d}$ in undoped WSe$_2$ FETs at a fixed gate voltage $V_{\rm g}$, extracted from the transfer characteristics shown in  Fig. \ref{fig:FIG3}(a). The observed increase in $I_{\rm d}$ with rising temperature confirms the semiconducting nature of pristine WSe$_2$. Figs.  \ref{fig:FIG3}(e) and (f) show the temperature dependence of $I_{\rm d}$ for W$_{0.83}$V$_{0.17}$Se$_2$ and W$_{0.68}$V$_{0.32}$Se$_2$ FETs respectively. In these doped samples, $I_{\rm d}$ decreases with increasing temperature, indicating a metal-like behavior. This suggests that the Fermi level in the doped samples moves closer to or even crosses the VBM, effectively transforming W$_{1-x}$V$_x$Se$_2$  into a degenerate semiconductor. This shift in Fermi energy, along with dopant-induced band structure modifications, is also supported by theoretical studies reported in the literature \cite{doping15}. This insulator-to-metal transition may also be partially attributed to the presence of the secondary metallic 1T-VSe$_2$ phase within the W$_{1-x}$V$_x$Se$_2$  matrix, as suggested by the Raman spectra. 
\begin{figure}
\centering
\includegraphics[width=8cm]{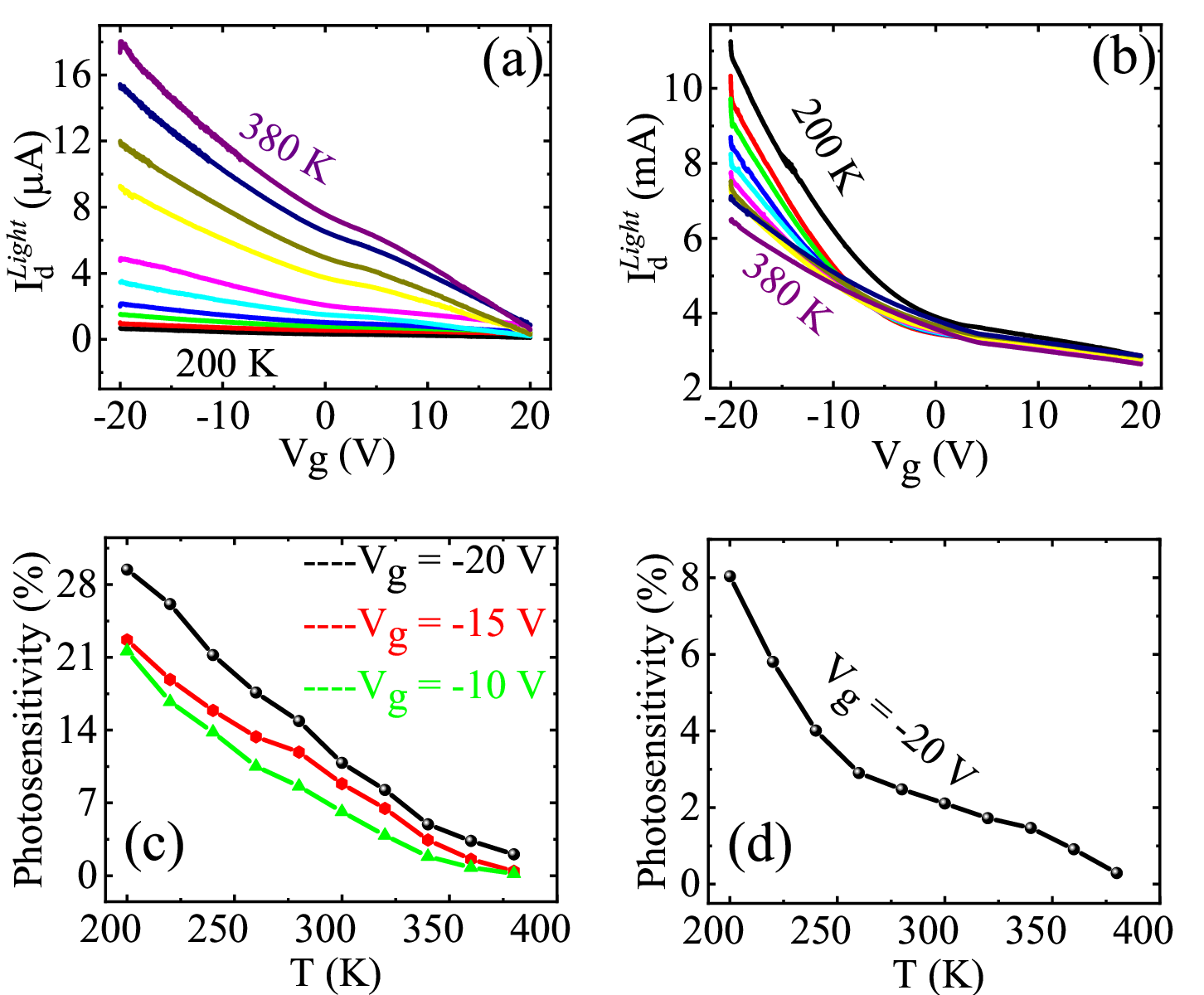}
\caption{(a) Temperature-dependent $I_{\rm d}$ - $V_{\rm g}$ characteristics of (a) pristine, (b) W$_{0.83}$V$_{0.17}$Se$_2$ FETs at $V_{\rm d}$ = 2 V under illumination with a 532 nm wavelength laser. (c) and (d) show the variation of photoconductive gain with temperature extracted from the curves in (a) and (b) respectively.}
\label{fig:FIG4}
\end{figure}
\par Figures \ref{fig:FIG4}(a) and (b) illustrate the temperature-dependent $I_{\rm d}$ vs $V_{\rm g}$ characteristics of pure WSe$_2$ and W$_{0.83}$V$_{0.17}$Se$_2$ FETs under illumination with a 532 nm wavelength laser with an aerial power density of 1.77 mW/$mm ^{2}$. The temperature-dependent behavior of $I_{\rm d}$ in light follows the same trend as in dark conditions; however, the magnitude of $I_{\rm d}$ increases under illumination. Figs. \ref{fig:FIG4}(c) and (d) present the photoconductive gain, calculated as: ${(I_{\rm d}^{\rm light}-I_{\rm d}^{\rm dark}}/I_{\rm d}^{\rm dark})\times 100\%$, where $I_{\rm d}^{\rm light}$ and $I_{\rm d}^{\rm dark}$ represent the drain current in the presence and absence of light, respectively. The photoconductive gain is observed to decrease with increasing temperature in both undoped and W$_{0.83}$V$_{0.17}$Se$_2$ FETs devices. This reduction can be attributed to enhanced electron-phonon interactions at elevated temperatures, which facilitate faster non-radiative recombination of photoexcited electron-hole pairs, thereby reducing the number of free carriers contributing to the photocurrent. A significant observation from Figs. \ref{fig:FIG4}(c) and (d) is the drastic reduction in photoconductive gain in the W$_{0.83}$V$_{0.17}$Se$_2$, and the almost negligible gain in the W$_{0.68}$V$_{0.32}$Se$_2$ device. Several factors may explain this trend: 1) Vanadium doping introduces mid-gap states within the bandgap of WSe$_2$ \cite{doping15}, which act as recombination centers for photogenerated electron-hole pairs, increasing non-radiative recombination, reducing the carrier lifetime, and consequently, the overall photocurrent. 2) V-doping contributes additional free hole carriers as confirmed by the FETs measurements, which screen the electric field generated by photogenerated electron-hole pairs, reducing the effectiveness of charge separation and transport, thus lowering the overall photocurrent. 3) In undoped WSe$_2$, a small number of photo carriers modulate the channel conductivity over an extended period, leading to high photoconductive gain. In doped WSe$_2$, the excess charge carriers neutralize this effect, causing a lower photo-gating-induced response and, hence, a weaker photocurrent. 4) Heavy doping levels in WSe$_2$ can modify the band structure, leading to Pauli blocking \cite{reason2}, where excess carriers fill available states near the valence band edges. This reduces the effective absorption of incident photons, leading to lower photo response. As a result, the W$_{0.68}$V$_{0.32}$Se$_2$ sample exhibits an almost zero photoconductive gain.
\begin{figure}
\centering
\includegraphics[width=8cm]{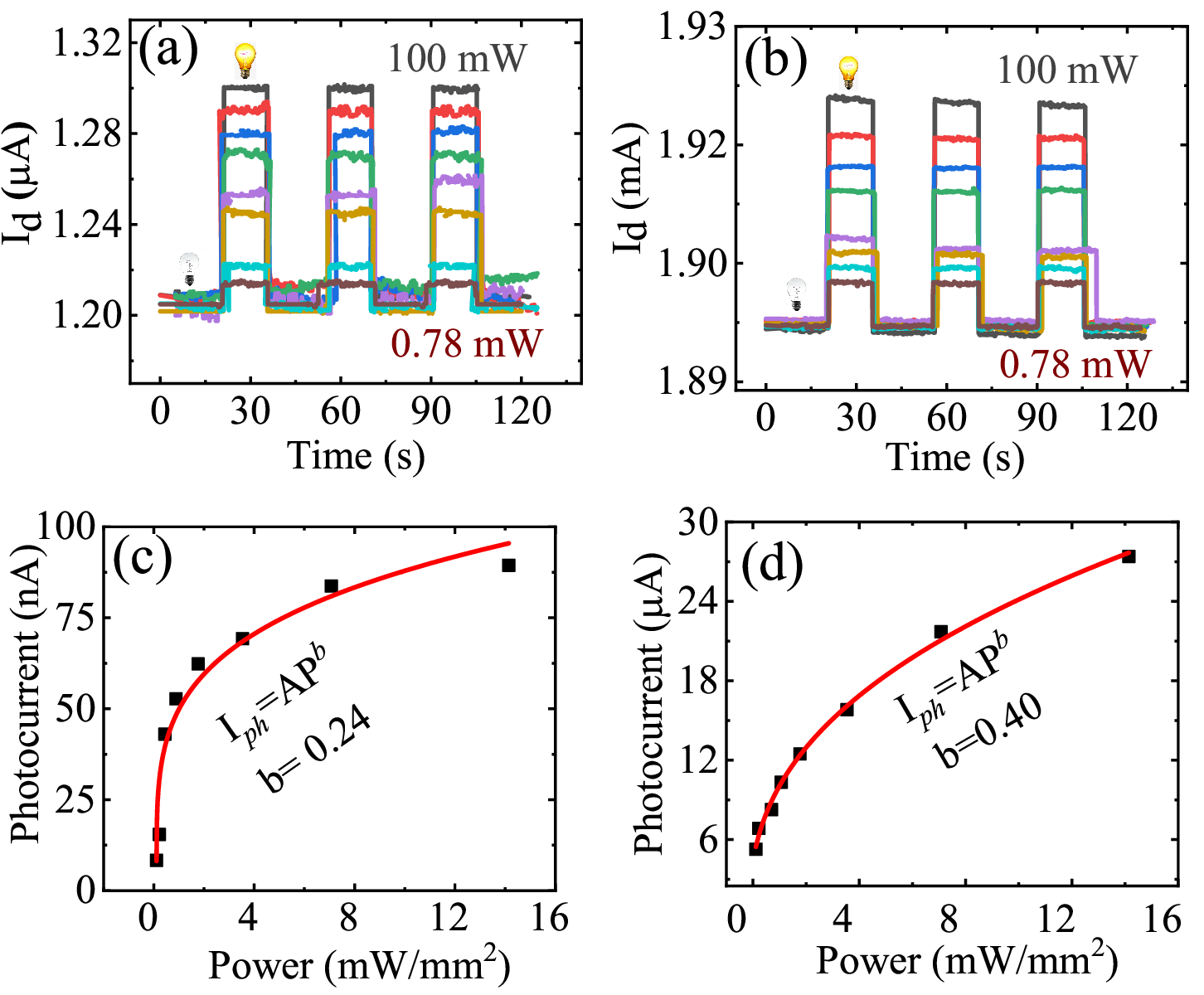}
\caption{ Time-resolved photo-response of $I_{\rm d}$ under illumination with a 532 nm wavelength laser at different power levels for (a) pristine and (b) W$_{0.83}$V$_{0.17}$Se$_2$ FETs at $V_{\rm g}$ = 0 V and $V_{\rm d}$ = 2 V at room temperature. (c) and (d) show the variation of photocurrent with laser power, extracted from the curves in (a) and (b) respectively.}
\label{fig:FIG5}
\end{figure}         
\par Figure \ref{fig:FIG5}(a) and (b) illustrate the time-resolved photo-response of $I_{\rm d}$ in both pure and W$_{0.83}$V$_{0.17}$Se$_2$ samples for a fixed $V_{\rm d}$ = 2 V and $V_{\rm g}$ = 0 V at room temperature. The samples were sequentially illuminated with a 532 nm wavelength laser at varying power levels (0.31, 1.56, 3.12, 6.25, 12.5, 25, 50 and 100 mW). Upon light exposure, $I_{\rm d}$  exhibits an instantaneous increase within a time resolution of approximately 50-60 ms, with no relaxation or slow tail. This suggests that the device exhibits a strong photocurrent response and consistent performance, thereby validating its capability for broadband photodetection. 
Figs. \ref{fig:FIG5}(c) and (d) illustrate the variation of photo current defined as $I_{\rm ph}$= $(I_{\rm d}^{\rm light}-I_{\rm d}^{\rm dark})$, with laser power, extracted from Figs. \ref{fig:FIG5}(a) and (b). As laser power increases, the incident photon flux rises, leading to a higher rate of electron-hole pair generation through photon absorption. This results in an increased carrier density, thereby enhancing the photocurrent. However, significant variations in photocurrent are observed at lower light intensities, while a saturation trend appears at higher intensities, indicating a power-law dependence of the form $I_{\rm ph}$= A$P ^{b} $ \cite{reason01,reason02} as depicted in Figs. \ref{fig:FIG5}(c) and (d), where A is a constant, P is the power in unit of mW/$mm ^{2}$, and b is the power indices 0.29 and 0.40 respectively. This saturation at higher intensities may be attributed to saturable absorption or optical limiting in the WSe$_2$ films, due to the Pauli-blocking effect \cite{reason2}. Additionally, local thermal effects could also contribute by enhancing electron-phonon scattering, thereby reducing carrier mobility. 
 
In conclusion, selenization of V$_2$O$_5$/WO$_3$ bilayer films present a promising route for scalable and controlled vanadium doping of WSe$_2$. Electrical transport measurements on W$_{1-x}$V$_x$Se$_2$ FETs reveal a significant enhancement in hole conduction, culminating in an insulator-to-metal transition that effectively makes it a degenerate p-type semiconductor on increasing the vanadium concentration. The observed reduction in photoconductive gain at the higher levels of substitution, attributable to enhanced carrier recombination and charge screening effects, provides valuable insights into the doping-induced band structure modifications and charge transport mechanisms in vanadium-doped WSe$_2$ films. The tunable conductivity and photoresponse of V-doped WSe$_2$ enable diverse optoelectronic applications, including high-performance p-type FETs, photodetectors with tunable sensitivity, and transparent electronics with enhanced optical stability. Its ability to suppress photogain under illumination  makes it ideal for ambient-light–resilient sensing and stable photoelectronic circuits,  positioning it as a candidate for next-generation optoelectronic applications with potential for wafer-scale integration.

This research has been performed at the United States Department of Defense funded Center of Excellence for Advanced Electro-photonics with 2D materials– Morgan State University, under grant No. W911NF2120213. The authors thank Mat Ivill for his assistance in interpreting the RBS spectra, as well as Frank Gardea and Owen Vail, the cooperative agreement managers of the Center, for their interest and support.

\clearpage
\section*{Supplementary Information}
\begin{center}
\textbf{Supplementary Material for Selenization on ``Selenization of V$_2$O$_5$/WO$_3$ Bilayers  for Tuned Optoelectronic Response of WSe$_2$ Films."}

\vspace{0.5em}

\end{center}

 
 \begin{figure*}
\centering
\includegraphics[width=17.5cm]{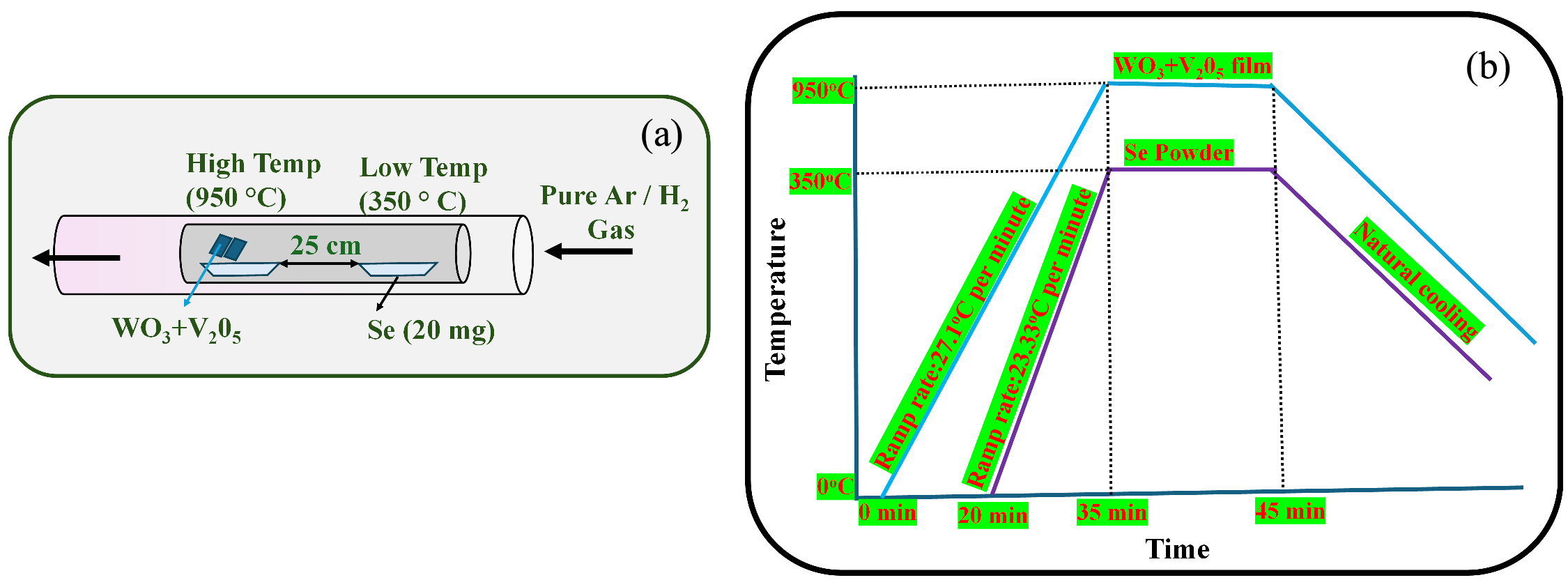}
\caption {(a) Schematic illustration of the selenization process of WO$_3$/V$_2$O$_5$ thin films in the atmospheric-pressure CVD reactor. (b) Ramp rate, dwell time, and dwell temperatures of the reaction.}
\label{fig:supp1}
\end{figure*}

\begin{figure*}
\centering
\includegraphics[width=17.5cm]{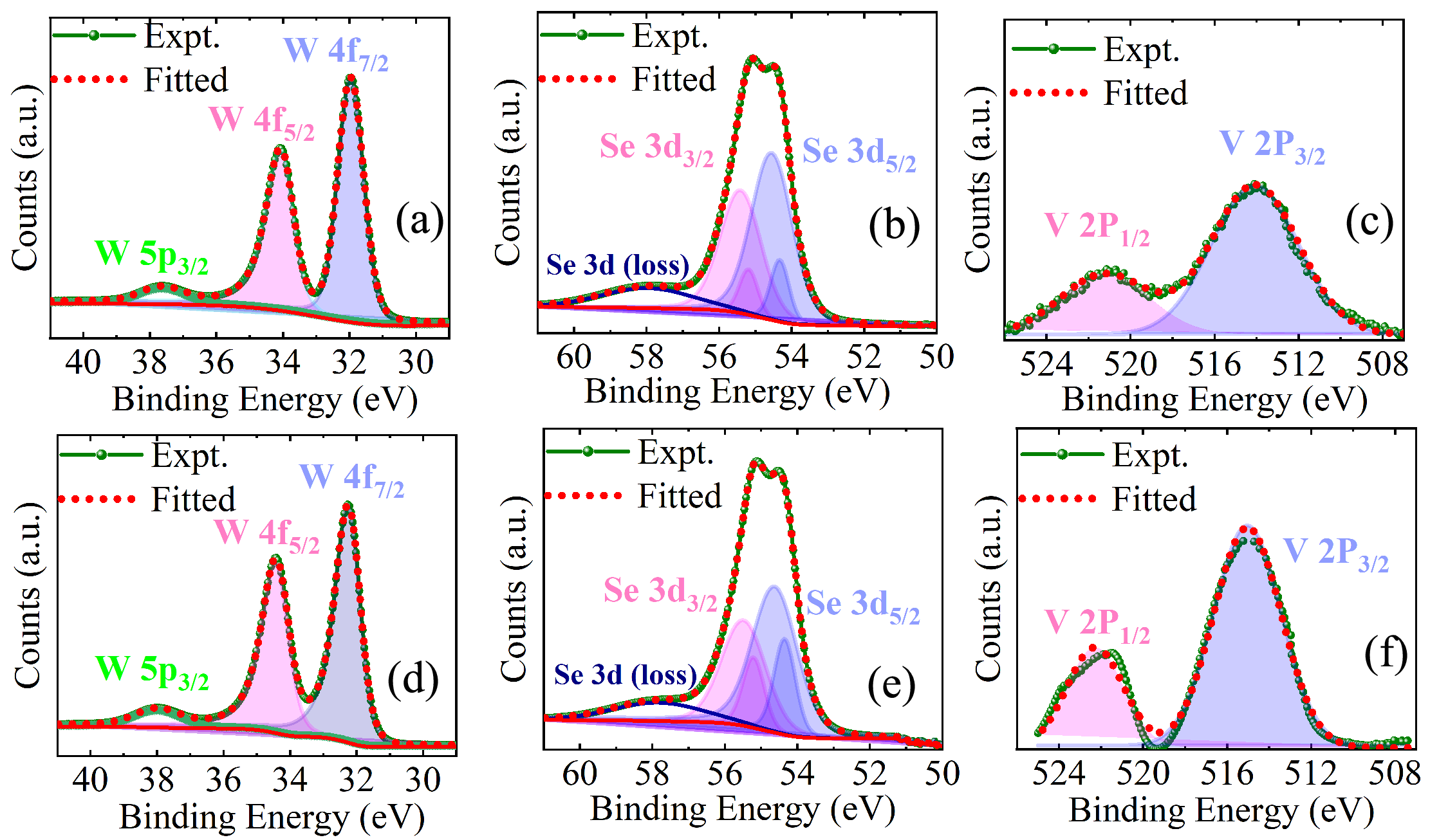}
\caption{(a),(b), and (c) show the XPS spectra of W 4f, Se 3d and V 2p of W$_{0.83}$V$_{0.17}$Se$_2$ sample,respectively. (d),(e) and (f) show the XPS spectra of W 4f, Se 3d and V 2p of W$_{0.68}$V$_{0.32}$Se$_2$ sample.}
\label{fig:supp2}
\end{figure*}

\begin{figure}
\centering
\includegraphics[width=8cm]{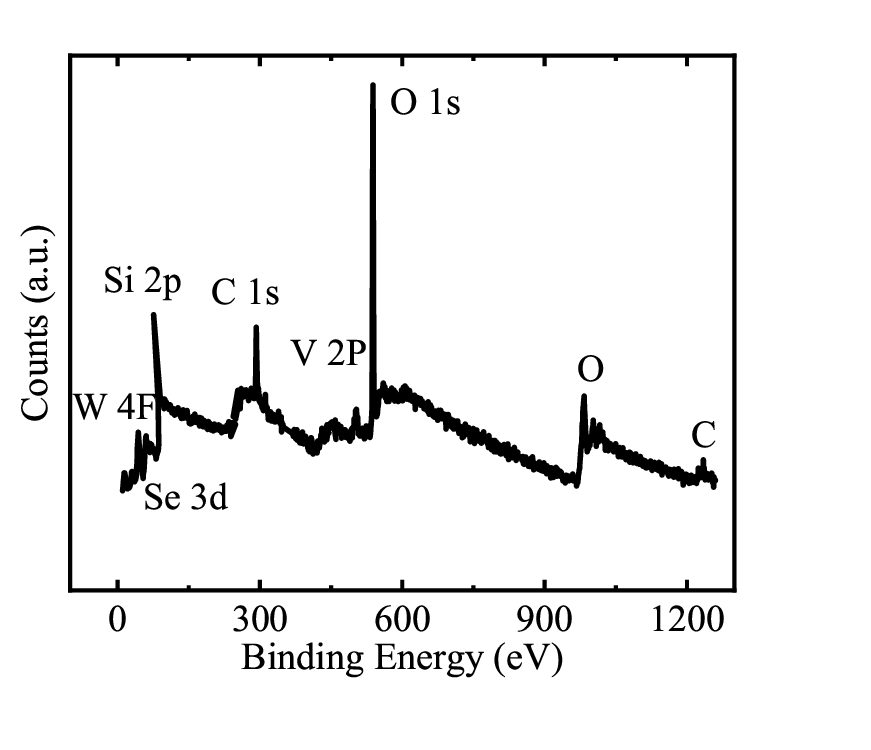}
\caption{Wide-scan XPS spectrum of our sample across the full binding energy range.
The observed  O 1s peak at around  538 eV is attributed to contributions from the SiO$_2$ substrate and chemically adsorbed oxygen bonded to tungsten. The other peaks are occur mainly from the contribution of C, Si, W, Se, and V. This indicates that there are  no other impurities present in our studied samples.}
\label{fig:supp3}
\end{figure}

This supplementary section presents the details of the procedures used for the thermal evaporation of  V$_2$O$_5$/WO$_3$ thin films, their subsequent selenization, and of the characterization techniques employed to analyse the structural characteristics of the W$_{1-x}$V$_x$Se$_2$ films.
Highly doped  silicon wafers with 300 nm thermally grown SiO$_2$ layer were thoroughly cleaned using acetone and isopropyl alcohol. First, ultra-thin films (1 nm to 3 nm) of V$_2$O$_5$ were deposited at the rate of ~ 0.1 Å/s followed by the growth of 10 nm thick WO$_3$ layer at the rate of ~ 0.3 Å/s without exposing the sample in ambient. The deposition was carried out in a vacuum chamber (base pressure ~1 $\times 10^{-6}$ torr) equipped with two thermal sources, utilizing high-purity V$_2$O$_5$ and WO$_3$ (99.99\%) as the source materials. High vapor pressures of V$_2$O$_5$ and WO$_3$  make it very easy to evaporate these oxides from a thermal source. The growth rate of each layer was monitored with a quartz crystal balance. The transformation of WO$_3$ into WSe$_2$ and WO$_3$/V$_2$O$_5$ into V-doped WSe$_2$ was carried out using the reactor, which features a two-zone tubular furnace with independent temperature control for each zone, and three mass flow controllers to regulate gas flow within the reactor. The films of V$_2$O$_5$ layer thickness of 0.0, 1.0, 3.0 nm are labeled as Sample A, B, and C respectively. 
\par A schematic representation of the substrate placement within the reactor is shown in Fig. \ref{fig:supp1}(a). The pure WO$_3$ film and WO$_3$/V$_2$O$_5$ bilayers on SiO$_2$/Si substrates were placed face-up in an alumina crucible boat in the high-temperature zone (Zone 1), while high-purity selenium powder (99.99\%) was positioned in a separate alumina crucible boat within the low-temperature zone (Zone 2), maintaining a 25 cm separation between the them. Before initiating the thermal cycle (as shown in Fig. \ref{fig:supp1}(b)), the reactor tube was evacuated to a pressure between  1 $\times 10^{-4}$ torr and 3 $\times 10^{-4}$ torr using a dry pump, followed by flushing with ultrahigh purity (99.9999\%) Ar/H$_2$ gas mixture (5\% H$_2$ in Ar) at a flow rate of 100 standard cubic centimeters per minute (SCCM) for 45 minutes. Throughout the reaction, the gas flow was maintained at 20 SCCM. The growth process involved heating Zone 1 to $950\,^\circ\mathrm{C}$ and Zone 2 to $350\,^\circ\mathrm{C}$, holding these temperatures constant for 10 minutes as illustrated in Fig. \ref{fig:supp1}(b). The Se vapors reduce WO$_3$, leading to the formation of intermediate WO$_x$Se$_y$ phases. The complete reaction results in the conversion of WO$_3$ into WSe$_2$, releasing volatile oxygen species (O$_2$, SeO$_2$): WO$_3$ + xSe → WSe$_2$ + SeO$_2$ (gas). Here, H$_2$ gas flow enhances reduction and improves crystallinity of the film. Upon completion of the selenization, the system was rapidly cooled to room temperature by opening the furnace lid.
\par After growth, films were characterized by Raman spectroscopy, Rutherford backscattering spectrometry (RBS), Atomic force microscopy (AFM), Conductive AFM (C-AFM), X-ray photoelectron spectroscopy (XPS), and  Scanning electron microscope (SEM).  Raman spectroscopy was performed at room temperature using a 532 nm excitation laser with a power of 2.5 mW. The measurements were conducted with an XploRA PLUS confocal Raman microscope (Horiba), ensuring precise spectral acquisition. All Raman peaks were calibrated against the well-defined 520.7 $cm ^{-1}$ optical phonon mode of silicon for accuracy. The RBS measurements were performed using a 2 MeV $He ^{+}$ beam from a National Electrostatics 5 SDH-2 positive ion accelerator, and the data interpreted using SIMNRA data analysis software \cite{RBS1}. AFM imaging was carried out using the Asylum Research MFP-3D Origin (Oxford Instruments) in tapping mode to analyze surface morphology and film uniformity. C-AFM data were collected using park system NX10 AFM system. The XPS data was acquired using a Phi Versa Probe II.  A monochromated Al k-alpha x-ray source with a photon energy of 1486.6 eV was used to generate the photoelectrons. The analysis area was 200\,{\textmu}m~$\times$~200\,{\textmu}m for all samples with a step size of 0.05 eV. The CasaXPS Version 2.3.26 was used for data analysis.
   
   \par To quantify the stoichiometry of the doped samples, RBS was utilized to determine the areal densities of the constituent elements. For the 1 nm V$_2$O$_5$/10 nm WO$_3$ sample, the areal densities were found to be $25 \times 10^{12}$~atoms/cm$^2$ for W, $5.2 \times 10^{12}$~atoms/cm$^2$ for V, and $57 \times 10^{12}$~atom/cm$^2$ for Se. Based on these values, the total metal atom density (W+V) is  $30.2 \times 10^{12}$~atoms/cm$^2$, yielding an atomic ratio of Se:(W+V) is 1.89:1 and a vanadium fraction of $x = 5.2/30.2 = 0.172$. Thus, the stoichiometric formula can be written as V$_{0.17}$W$_{0.83}$Se$_2$. Similarly, for the 3 nm V$_2$O$_5$/10 nm WO$_3$ sample sample, the RBS analysis determined areal densities of $20 \times 10^{12}$~atoms/cm$^2$ for W, $9.2 \times 10^{12}$~atoms/cm$^2$ for V, and $55.5 \times 10^{12}$~atoms/cm$^2$ for Se. The total metal atom density (W+V) is  $29.2 \times 10^{12}$~atoms/cm$^2$, resulting in a Se:(W+V) ratio of 1.90:1 and $x = 9.2/29.2 = 0.315$. Accordingly, the formula is expressed as V$_{0.32}$W$_{0.68}$Se$_2$. 

\par Table \ref{tab2} summarizes the chemical compositions of Sample B, and Sample C, as calculated from RBS analysis.

\begin{table}[!htbp]
\centering
\caption{Atomic percentage of samples from RBS.}
\label{tab2}
\begin{tabular}{|c|c|c|}
\hline
Element (at.\%) & Sample B & Sample C \\
\hline
W & 28.669 & 23.613 \\
Se &  65.368 & 65.525 \\
V  & 5.963 & 10.862 \\
Stoichiometry & V$_{0.17}$W$_{0.83}$Se$_2$ & V$_{0.32}$W$_{0.68}$Se$_2$ \\
\hline
\end{tabular}
\end{table} 
  

\par Figure \ref{fig:supp2} shows the high resolution XPS of the W 4f, V 2p, and Se 3d, fitted following the approach presented in the literature \cite{XPS0}. In the W$_{0.83}$V$_{0.17}$Se$_2$ sample, Fig. \ref{fig:supp2}(a) presents the W 4f and 5p spectrum, the 5p peak is at 37.63 eV and the 4f doublet peaks observed at 31.88 and 34.04 eV, corresponding to the W 4f$_7$$_/$$_2$ and W 4f$_5$$_/$$_2$ binding energies, respectively. Fig. \ref{fig:supp2}(b) displays the Se 3d spectrum, where the broad peak centered at 57.51 eV is attributed to an Se 3d loss peak and the peaks at 54.56 and 55.42 eV are attributed to Se 3d$_5$$_/$$_2$ and Se 3d$_3$$_/$$_2$ binding energies, respectively. The another set of peak of Se 3d$_5$$_/$$_2$ and Se 3d$_3$$_/$$_2$ binding energies at around 54.34 eV and 55.20 eV which may be associated with the formation of a secondary metallic VSe$_2$ phase \cite{XPS1}. A four-component fitting is more appropriate in this case than a two-component fitting due to the atypical asymmetry in the Se 3d doublet shown in the experimental data. Typically, the asymmetry in the Se 3d doublet peak reflects an area ratio of 6:4 electrons between the 5/2 and 3/2 states, respectively. However, our XPS data seemingly shows the 3/2 peak appearing more intense than the 5/2 peak, which is problematic, as this would contradict the expected quantum mechanical spin-orbit splitting ratio.
\par Figure \ref{fig:supp2}(c) shows the V 2p spectrum, with the peaks at 514 and 521.2 eV correspond to V 2p$_3$$_/$$_2$ and V 2p$_1$$_/$$_2$ binding energies, respectively. XPS analysis also demonstrates that in the W$_{0.68}$V$_{0.32}$Se$_2$ sample, shown in Fig. \ref{fig:supp2}(d), the W 4f peaks exhibit a noticeable shift towards higher binding energies by approximately 0.32 eV  compared to the W$_{0.83}$V$_{0.17}$Se$_2$ sample. This shift in binding energies may be attributed to the formation of a secondary metallic VSe$_2$ phase, which locally doped the WSe$_2$. This interpretation is supported by the observed shift in the Fermi level toward the valence band maximum (VBM) of WSe$_2$, as reported in the literature \cite{doping017}. It is also consistent with our optoelectronic measurements, which show a significant increase in drain current accompanied by a reduction in photocurrent gain. 
The chemical composition, as detailed in Table \ref{tab}, extracted from XPS data indicates that the atomic percent of V was 6.2\% and 3.3\%, respectively. The doubling of the V content of Sample C from Sample B  is in agreement with the expected and RBS measurements. We attribute the lower V percent from XPS as being due to the lower kinetic energy of the V 2p photoelectrons versus the W 4f and Se 3d. Attempts to remove the adventitious carbon layer through sputtering led to significant preferential sputtering of selenium.

\begin{table}[!htbp]
\centering
\caption{Chemical composition of the samples.}
\label{tab}
\begin{tabular}{|c|c|c|}
\hline
Element (at.\%) & Sample B & Sample C \\
\hline
W & 35.0486 & 33.585 \\
Se &  61.6524 & 60.2017 \\
V  & 3.299 & 6.21332\\
\hline
\end{tabular}
\end{table} 

\par
\begin{figure*}[!htbp]
\centering
\includegraphics[width=17.5cm]{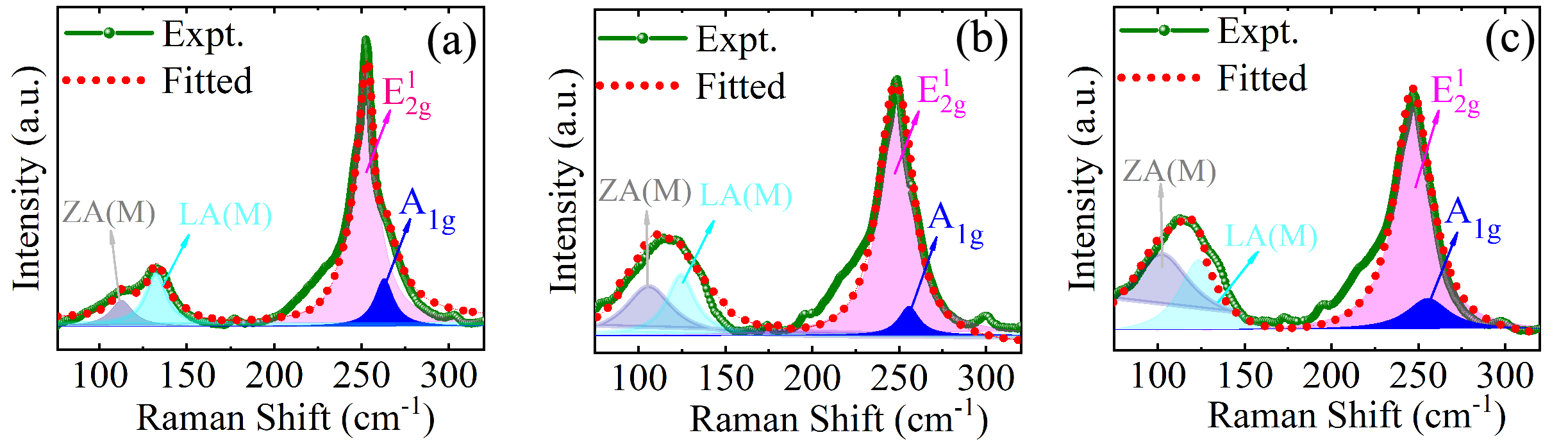}
\caption{(a), (b), and (c) represent the raman spectra of pure WSe$_2$, W$_{0.83}$V$_{0.17}$Se$_2$, and W$_{0.68}$V$_{0.32}$Se$_2$, respectively. The green dots represent the experimental raman data, while the red dashed line denotes the cumulative Lorentzian fit. Individual vibrational modes of the WSe$_2$ film are deconvoluted and shown as distinct peaks in different colors.}
\label{fig:supp4}
\end{figure*}

 Figure \ref{fig:supp4} presents the detailed Lorentzian fitting of  WSe$_2$, W$_{0.83}$V$_{0.17}$Se$_2$, and W$_{0.68}$V$_{0.32}$Se$_2$ samples, highlighting the shift of characteristic raman peaks and the enhancement of defect-activated modes with increasing vanadium concentration. Table \ref{tab1} summarizes the fitting parameters such peak position, area under the curve, and Full width half maxima (FWHM), extracted from these spectra by following  the method described in literature \cite{Raman}. These observations are consistent with trends reported in the literature \cite{doping015},\cite{Raman0}. 
\par  During the initial Raman fitting, a weak shoulder near 220 $cm ^{-1}$ was consistently observed across all spectra. Although this region was initially included to enhance the fit quality, but excluded from the final presented results due to the absence of any confirmed Raman active modes, either in pristine or doped WSe$_2$ at this position. This feature is most likely attributable to second-order or zone-edge phonon scattering processes under resonant excitation with 532 nm laser, or minor disorder-activated contributions, rather than a fundamental vibrational mode.
To maintain physical consistency and avoid over fitting with unassigned peaks, the 220 $cm ^{-1}$ region was excluded from the final fits, resulting in a slightly reduced coefficient of determination ($R ^{2}$). It ensures that only physically justified peaks are shown in Fig \ref{fig:supp4}.

\begin{table*} [!htbp]
\centering
\caption{Raman peak parameters for Pure WSe$_2$, W$_{0.83}$V$_{0.17}$Se$_2$, and W$_{0.68}$V$_{0.32}$Se$_2$}
\resizebox{16cm}{!}{%
\begin{tabular}{|l|cccc|cccc|cccc|}
\hline
\textbf{} & \multicolumn{4}{c|}{\textbf{Peak Position}} & \multicolumn{4}{c|}{\textbf{Area}} & \multicolumn{4}{c|}{\textbf{FWHM($cm ^{-1}$)}} \\
\textbf{} & $E^{\rm 1}_{\rm 2g} $ & $A _{\rm 1g} $ & ZA(M) & LA(M) & $E^{\rm 1}_{\rm 2g} $ & $A _{\rm 1g} $ & ZA(M) & LA(M) & $E^{\rm 1}_{\rm 2g} $ & $A _{\rm 1g} $ & ZA(M) & LA(M) \\
\hline
Pure WSe$_2$ & 253.09 & 262.87 & 112.63 & 134.11 & 71691 & 14808 & 7567 & 15471 & 16.21 & 13.49 & 15.90 & 19.44 \\
W$_{0.83}$V$_{0.17}$Se$_2$ & 247.60 & 255.65 & 105.88 & 124.5 & 67106 & 12092 & 30407 & 15670 & 22.29 & 14.35 & 33.25 & 20.21 \\
W$_{0.68}$V$_{0.32}$Se$_2$ & 247.14 & 255.41 & 102.18 & 121.24 & 66292 & 11075 & 35737 & 25469 & 24.09 & 29.58 & 41.08 & 28.13 \\
\hline
\end{tabular}}
\label{tab1}
\end{table*}

\end{document}